\documentclass[prd,aps,showpacs,tightenlines]{revtex4}  
\usepackage{mathrsfs}
\usepackage{amsmath}
\usepackage{amssymb}
\usepackage{epsfig}
\usepackage{graphicx}
\usepackage{booktabs}
\usepackage{multirow}
\usepackage{subfigure}

\begin{document}
\newcommand{\psl}{ p \hspace{-1.8truemm}/ }
\newcommand{\nsl}{ n \hspace{-2.2truemm}/ }
\newcommand{\vsl}{ v \hspace{-2.2truemm}/ }
\newcommand{\epsl}{\epsilon \hspace{-1.8truemm}/\,  }


\title{   $S$-wave $K\pi$ contributions to the hadronic charmonium  $B$ decays in the perturbative QCD approach }
\author{Zhou Rui$^1$}\email{jindui1127@126.com}
\author{Wen-Fei Wang$^2$}\email{wfwang@sxu.edu.cn}
\affiliation{$^1$College of Sciences, North China University of Science and Technology, Tangshan 063009,  China}
\affiliation{$^2$Institute of Theoretical Physics, Shanxi University, Taiyuan, Shanxi 030006, China}

\date{\today}

\begin{abstract}
We extend our recent works on  the two-pion $S$-wave resonance contributions to the kaon-pion ones
in the $B$ meson hadronic charmonium decay modes based on the perturbative QCD approach.
The $S$-wave $K\pi$ timelike form factor in its distribution amplitudes is described by the LASS parametrization,
which consists of the $K^*_0(1430)$ resonant state together with an effective range nonresonant component.
The predictions for the decays $B\rightarrow J/\psi K\pi$ in this work agree well with the experimental results
from the  $BABAR$ and Belle collaborations. We also discuss theoretical uncertainties,
indicating  that the results of this work, which can be tested by the LHCb and Belle-II experiments, are reasonably accurate.
\end{abstract}


\maketitle

\section{Introduction}
Within the quasi-two-body approximation, the $B$ meson decays to the hadronic three-body final states
can be restricted to the specific kinematical configurations, in which
the three daughter mesons are quasialigned in the rest frame of the $B$ meson.
The related processes  can be denoted as $B\rightarrow M_1(M_2M_3)$, where $M_1$ is the bachelor particle,
and  the remaining $M_2M_3$ pair   proceeds via intermediate resonant states.
The final state interactions are expected to be suppressed in such conditions.
According to the kinematics and angular momentum conservation,  the resonant states are predominantly
found in the scalar ($S$-wave), vector ($P$-wave), or tensor ($D$-wave) meson spectrum, etc.
Studies of the quasi-two-body decays will help us to clarify the nature of the resonances involved.
The final state phase space can be represented in a Dalitz plot (DP),  which provides  information about  the weak
and strong phases in the decay processes.

Among the numerous three-body $B$ decay channels, the category including a vector charmonium state and
 one kaon-pion pair via $b\rightarrow s c\bar{c}$ and $b\rightarrow d c\bar{c}$ transitions is particularly interesting.
For instance, the interference between the $S$- and $P$-waves of the $K\pi$ systems produced in $B^0\rightarrow J/\psi K\pi$
decays \cite{prd71032005}
allows one to resolve the sign ambiguity of $\cos(\beta)$, where   $\beta$ is the Cabibbo-Kobayashi-Maskawa (CKM) phase.
These decays have also been regarded as a source of information about the composition of the scalar-meson-like $\kappa$,
which may exist as   four-quark states in the low invariant mass regions \cite{prd15267}.
Of course, the $K\pi$ pair is not the only resonance source. In recent years,
 many charmoniumlike  resonant structures (a minimal quark content  must be the exotic combination $c\bar{c}u\bar{d}$),
such as $Z^+(4430)$, $Z^+(4050)$, $Z^+(4250)$, 
have also been observed in the $\psi\pi$ [$\psi=J/\psi, \psi(2S)$] invariant mass spectrum in the $B\rightarrow \psi K\pi$ decays \cite{prl100142001,prd79112001,prd78072004,prd80031104,prd85052003,prd90112009},
 which are not easy to   accommodate  in the quark model of hadrons \cite{pr429243,pr4541}.
 DP analyses of such processes provide opportunities for the studies of the spectroscopy of these new structures.

Using the sample corresponds to an integrated luminosity of  $413 \text{fb}^{-1}$,
the $BABAR$ Collaboration studied the resonant structures in the $B\rightarrow J/\psi K\pi$
and $B\rightarrow \psi(2S) K\pi$ decays in~Ref. \cite{prd79112001}.
The corresponding DP analyses in the two channels show   important contributions in the $K\pi$ $S$-wave systems.
Subsequently, the Belle Collaboration revealed a rich resonance spectrum in the $K\pi$ mass distribution
based on a $711 \text{fb}^{-1}$ data sample collected  at the asymmetric-energy $e^+e^-$ collider KEKB \cite{prd88074026,prd90112009}.
They found  clear evidence for the $K^*_0(1430)$ resonance with a $22.0\sigma$ significance for the decay modes including $J/\psi$,
but only a $1.6\sigma$ signal for that of $\psi(2S)$ analogues.
Furthermore, the $K^*_0(1430)$  fit fraction in $B\rightarrow \psi(2S) K\pi$
was  comparable 
with its previous  measurements \cite{prd80031104} and the  LHCb's data~\cite{prl112222002}.
Recently, the $B_s\rightarrow \psi(2S)K\pi$ decay mode was first observed by the LHCb Collaboration;
the fit fractions of the $S$-wave component reach  $0.339\pm 0.052$, with   statistical uncertainty only~\cite{plb747484}.

On the theoretical side,
several approaches have been used for describing charmless three-body $B$ decays involving   $K\pi$ systems.
For example, in   Refs.~\cite{prd76094006,prd88114014,prd94084015,prd89094007,sc58031001}, the authors predicted the branching ratios and direct $CP$ violations in charmless three-body decays $B\rightarrow K\pi\pi$ and $B\rightarrow KK\pi$ using a model based on the factorization approach. The method was extended further in charmless three-body $B_s$ decays in  
Ref.~\cite{prd89074025}.   In Ref. \cite{prd79094005}, the $CP$ violation and  the contributions of the strong kaon-pion
interactions have been studied in  $B\rightarrow K \pi\pi$ decays using an approximate construction of relevant scalar
and vector form factors. The $K^*$ resonance effects  on direct $CP$ violation have been taken into account based on
the QCD factorization scheme \cite{prd81094033},
while, the three-body $B$ meson decays with the charmonium mesons in the final states have not
received much attention  in the literature.

In our previous works, the decays $B_{(s)}\rightarrow (J/\psi,\eta_c) \pi\pi$ \cite{prd91094024,epjc76675},
as well as the corresponding $\psi(2S), \eta_c(2S)$ modes \cite{170802869,epjc77199,ma-cpc2017},
with the pion pair in $S$-wave resonant states, have been studied in the perturbative QCD (PQCD)~\cite{ppnp5185,prd63-074009,Wang-2016} framework
by introducing two-pion distribution amplitudes for  the resonances \cite{plb561258,Wang-2014a}.
These processes have been well described by a series of scalar resonances
such as $f_0(500)$, $f_0(980)$, $f_0(1500)$,  $f_0(1790)$, and so on.
In the present paper,  motivated by the recent detailed DP analyses of the $K\pi$ invariant mass spectrum
by the $BABAR$ \cite{prd79112001}, Belle \cite{prd90112009,prd88074026}, and LHCb \cite{prl112222002} collaborations,
we will work on the decays of $B_{(s)}\rightarrow (J/\psi,\psi(2S))K\pi$,
and we will focus on the $K\pi$ pair originating from a scalar quark-antiquark state,
while other partial wave and charmoniumlike resonances are beyond the scope of the present analysis.
The $S$-wave contributions are parametrized into the timelike scalar form factors
involved in the kaon-pion distribution amplitudes.
For these form factors, we will adopt the LASS parametrization in   Ref.~\cite{npb296493},
which consists of a linear combination of the $K^*_0(1430)$ resonance and a nonresonant term coming from elastic scattering.
By introducing the kaon-pion distribution amplitudes, the $S$-wave contributions of the related three-body $B$ decays
can be simplified into the quasi-two-body processes $B\rightarrow \psi (K\pi)_{\text{$S$-wave}}\rightarrow \psi K\pi$.
Following the steps of Refs. \cite{prd91094024,epjc77199}, the decay amplitude of $B \rightarrow \psi (K\pi)_{\text{$S$-wave}}$
can be written as the convolution
\begin{eqnarray}
\mathcal{A}(B \rightarrow \psi (K\pi)_{\text{$S$-wave}})=\Phi_B \otimes H\otimes \Phi^{\text{$S$-wave}}_{K\pi} \otimes \Phi_{\psi},
\end{eqnarray}
where the hard kernel $H$ includes the leading-order contributions plus next-to-leading-order (NLO) vertex corrections.
The $B$ meson (charmonium, $S$-wave $K\pi$ pair) distribution amplitude $\phi_B$ ($\phi_{\psi}, \phi^{\text{$S$-wave}}_{K\pi}$) absorbs the nonperturbative dynamics in the  hadronization processes.

The layout of this paper is as follows. In Sec. \ref{sec:framework},
elementary kinematics, meson distribution amplitudes, and the required timelike scalar form factor are described.
In Sec. \ref{sec:results}, we present a discussion following the presentation of the significant results on branching ratios.
 Finally, Sec. \ref{sec:sum} will be the conclusion of this work.

\section{ framework}\label{sec:framework}

\begin{figure}[!htbh]
\begin{center}
\vspace{1.5cm} \centerline{\epsfxsize=7 cm \epsffile{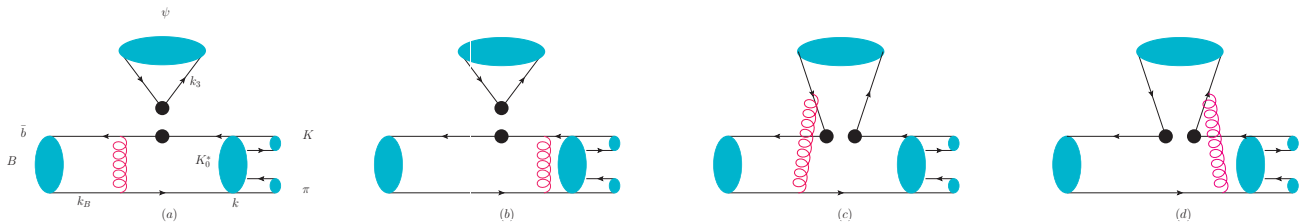}}
\caption{The leading-order Feynman diagrams for the
quasi-two-body decays  $B\rightarrow \psi K_0^* \rightarrow \psi K\pi$. The first two are factorizable
and the last two are nonfactorizable, and $K_0^*$ is the $S$-wave intermediate state.}
 \label{fig:femy}
\end{center}
\end{figure}

In the light-cone coordinates,  the kinematic variables  of the decay $B (p_B)\rightarrow \psi (p_3) (K\pi)(p)$
can be described in the $B$ meson rest frame as
\begin{eqnarray}
 p_B&=&\frac{M}{\sqrt{2}}(1,1,\textbf{0}_{T}),\quad p_3=\frac{M}{\sqrt{2}}(r^2,1-\eta,\textbf{0}_{T}),\quad  p=\frac{M}{\sqrt{2}}(1-r^2,\eta,\textbf{0}_{T}),
\end{eqnarray}
with the mass ratio $r=m/M$, and where $m(M)$ is the mass of the charmonium ($B$) meson,
the variable $\eta=\omega^2/(M^2-m^2)$,  and the invariant mass squared $\omega^2=p^2$
for the kaon-pion pair.  As usual we also define the kaon momentum  $p_1$ and pion momentum $p_2$ as
\begin{eqnarray}
 p_1=\frac{M}{\sqrt{2}}((1-r^2)\xi, \eta (1-\xi),\textbf{p}_{1T}),\quad
 p_2=\frac{M}{\sqrt{2}}((1-r^2)(1-\xi), \eta \xi,\textbf{p}_{2T})
\end{eqnarray}
with $\xi$ being the kaon momentum fraction. The momenta satisfy the momentum conservation $p=p_1+p_2$.
The three-momenta of the kaon and    charmonium in the $K\pi$ center of mass  are given by
\begin{eqnarray}
|\vec{p}_1|=\frac{\sqrt{\lambda(\omega^2,m_K^2,m_{\pi}^2)}}{2\omega}, \quad
|\vec{p}_3|=\frac{\sqrt{\lambda(M^2,m^2,\omega^2)}}{2\omega},
\end{eqnarray}
respectively, with $m_K$ ($m_{\pi})$ the kaon (pion) mass and  the K$\ddot{a}$ll$\acute{e}$n function $\lambda (a,b,c)= a^2+b^2+c^2-2(ab+ac+bc)$.
For the valence quarks, momenta   $k_B,k_3,k$, whose notations are displayed in Fig. \ref{fig:femy}, are chosen as
\begin{eqnarray}
  k_B&=&(0,\frac{M}{\sqrt{2}}x_B,\textbf{k}_{BT}),\quad k_3=(\frac{M}{\sqrt{2}}r^2x_3,\frac{M}{\sqrt{2}}(1-\eta)x_3,\textbf{k}_{3T}),\quad  k=(\frac{M}{\sqrt{2}}z(1-r^2),0,\textbf{k}_{T}),
\end{eqnarray}
where $k_{iT}$, $x_i$ represent the transverse momentum and longitudinal
momentum fraction of the quark inside the meson, respectively.

The $B$ meson can be treated as a heavy-light system, whose wave function in impact coordinate space can be expressed by \cite{ppnp5185}
\begin{eqnarray}
\Phi_{B}(x,b)=\frac{i}{\sqrt{2N_c}}[(\rlap{/}{p_B}+M)\gamma_5\phi_{B}(x,b)],
\end{eqnarray}
where $b$ is the conjugate variable of the transverse momentum of the valence quark of the meson,
and $N_c$ is the color factor. The distribution amplitude $\phi_{B}(x,b)$ is adopted in  the same form as it was in Refs. \cite{ppnp5185,prd65014007}
\begin{eqnarray}
\phi_{B}(x,b)=N x^2(1-x)^2\exp[-\frac{x^2M^2}{2\omega^2_b}-\frac{\omega^2_bb^2}{2}],
\end{eqnarray}
with  shape parameter $\omega_b=0.40\pm 0.04$ GeV for the $B_{u,d}$
mesons and $\omega_b=0.50\pm 0.05$ GeV for the $B_s$ meson.
The normalization constant $N$ is related to the decay constant $f_{B}$ through
\begin{eqnarray}
\int_0^1\phi_{B}(x,b=0)d x=\frac{f_{B}}{2\sqrt{2N_c}}.
\end{eqnarray}
For the considered decays, the vector charmonium meson
is longitudinally polarized. The longitudinal polarized component
of the wave function is defined as \cite{prd90114030,epjc75293}
\begin{eqnarray}
\Phi_{\psi}^L=\frac{1}{2\sqrt{N_c}}[m \rlap{/}{\epsilon_L}  \phi^L (x,b)+\rlap{/}{\epsilon_L} \rlap{/}{p_3}\phi^t (x,b)],
\end{eqnarray}
with the longitudinal polarization vector $\epsilon_L=\frac{1}{\sqrt{2}r}(-r^2,1-\eta, \textbf{0}_{T})$.
For the twist-2 (twist-3) distribution amplitudes $\phi^L (\phi^t)$,
the same form and parameters are adopted as in Refs. \cite{prd90114030,epjc75293}.

The $S$-wave kaon-pion distribution amplitudes are introduced
in analogy with the case of two-pion ones \cite{prd91094024,plb561258},
 which are organized into
\begin{eqnarray}\label{eq:fuliye2}
\Phi_{K\pi}^{\text{$S$-wave}}=\frac{1}{\sqrt{2N_c}}[\rlap{/}{p}\phi^{I=1/2}_{v\mu=-}(z,\xi,\omega^2)+
\omega\phi_s^{I=1/2}(z,\xi,\omega^2)+\omega(\rlap{/}{n}\rlap{/}{v}-1)\phi^{I=1/2}_{t\mu=+}(z,\xi,\omega^2)],
\end{eqnarray}
where $n=(1,0,\textbf{0}_{T})$  and $v=(0,1,\textbf{0}_{T})$ are two dimensionless vectors.
For $I=\frac{1}{2}$, $\phi^{I=1/2}_{v\mu=-}$ contributes at twist-2, while $\phi_s^{I=1/2}$ and $\phi^{I=1/2}_{t\mu=+}$
contribute at twist-3.
It is worthwhile to stress that  this kaon-pion system
has   similar asymptotic   distribution amplitudes (DAs) as the ones for a light scalar meson \cite{plb730336},
but we replace the scalar decay constants with the timelike form factor:
\begin{eqnarray}\label{eq:phi0st}
\phi^{I=1/2}_{v\mu=-}(z,\xi,\omega^2)&=&\phi^0=\frac{3}{\sqrt{2N_c}}F_s(\omega^2)z(1-z)
[\frac{1}{\mu_S}+B_13(2z-1)],
\nonumber\\
\phi_s^{I=1/2}(z,\xi,\omega^2)&=&\phi^s=\frac{1}{2\sqrt{2N_c}}F_s(\omega^2),\nonumber\\
 \phi^{I=1/2}_{t\mu=+}(z,\xi,\omega^2)&=&\phi^t=\frac{1}{2\sqrt{2N_c}}F_s(\omega^2)(1-2z),
\end{eqnarray}
where $\mu_S=\frac{m_S}{m_2-m_1}$ and $m_S$ and $m_{1,2}$ are the scalar meson mass and 
  running current quark masses, respectively; their values can be found in Refs. \cite{prd73014017,prd77014034,prd78014006}.
 $B_{1}$ is the first odd Gegenbauer moment for the light scalar mesons.
According to Refs. \cite{prd73014017,prd77014034,prd74114010}, there are two scenarios for the scalar mesons.
In scenarios I, all scalar mesons are viewed as the conventional  two-quark states. 
In scenarios II, the light scalar mesons below or near 1 GeV are treated as the  four-quark states, 
 while those above 1 GeV scalar mesons such as $f_0(1370)$, $K_0^*(1430)$, $a_0(1450)$, and so on
 are regarded as the ground states of $q\bar{q}$.
As noticed,   scenario II is more favored for explaining the $B^+\rightarrow K_0^*(1430)\pi^+ $ data measured
by both $BaBar$ \cite{prd78012004} and Belle \cite{prl96251803}. Besides,
scenario II is also supported by a lattice calculation \cite{npb578367} and the recent Regge trajectory calculation \cite{epjc77431}.
Hence, we prefer to use the Gegenbauer moments $B_{1}=-0.57\pm 0.13$ at the 1 GeV scale in scenario II obtained using the QCD sum rule method \cite{prd73014017,prd77014034}.

As is known, the relativity Breit-Wigner (RBW) model is unsuitable for describing
the $K\pi$ $S$-wave contributions because the  broad $\kappa$ and $K^*_0(1430)$ resonance
interferes strongly with a slowly varying nonresonant (NR) component.
Detailed discussions of the $S$-wave $K\pi$ systems in the isobar model,  $K$-matrix model, and
model-independent partial wave analysis method can be found in Refs.~\cite{prl89-121801,prd73-032004,plb653-1,plb681-14}.
In this work, we parametrize the timelike scalar form factor $F_s(\omega^2)$ for the $S$-wave $K\pi$ systems
 by the LASS line shape \cite{npb296493}, which has been widely adopted in the experimental data analysis,
its expression is given as \cite{prd72072003,prd79112001}
\begin{eqnarray}\label{eq:lass}
F_s(\omega^2)=\frac{\omega}{|\vec{p}_1|[\cot(\delta_B)-i]}+e^{2i\delta_B}\frac{m_0^2
\Gamma_0/|\vec{p}_{10}|}{m_0^2-\omega^2-im_0^2 \frac{\Gamma_0}{\omega }\frac{|\vec{p}_1|}{|\vec{p}_{10}|}}\;,
\end{eqnarray}
where the first term is an empirical term from inelastic scattering and the second term is
the resonant contribution with a phase factor to retain unitarity.
  $m_0=1.435$ GeV and $\Gamma_0=0.279$ GeV  \cite{prd79112001} are the pole mass and width of
the $K^*_0(1430)$ resonance state.  $|\vec{p}_{10}|$  is $|\vec{p}_{1}|$ evaluated at the $K\pi$ pole mass.
The   phase factor $\cot(\delta_B)$ is defined as
\begin{eqnarray}\label{eq:ar}
\cot(\delta_B)=\frac{1}{a|\vec{p}_1|}+\frac{r|\vec{p}_1|}{2},
\end{eqnarray}
with the shape parameters $a=1.94$ and $r=1.76$ \cite{prd79112001}.

The differential branching ratio for the $B\rightarrow \psi (K\pi)_{S-wave}$ decay takes the explicit form
\begin{eqnarray}\label{eq:dfenzhibi}
\frac{d \mathcal{B}}{d \omega}=\frac{\tau \omega|\vec{p}_1||\vec{p}_3|}{32\pi^3M^3}|\mathcal{A}|^2.
\end{eqnarray}
Since the $S$-wave kaon-pion distribution amplitude in~Eq.~(\ref{eq:fuliye2})
has the  same Lorentz structure as that of two-pion ones in Ref. \cite{prd91094024},
the decay amplitude $\mathcal{A}$ here can be straightforwardly obtained just by
replacing the twist-2 or twist-3 DAs  of the $\pi\pi$ system with
the corresponding  twists of the $K\pi$ one in Eq. (\ref{eq:phi0st}).
In addition, we also consider the NLO vertex corrections to the factorizable diagrams in Fig. \ref{fig:femy},
whose effects are included by the  modifications to the  Wilson coefficients as usual \cite{bbns1,bbns2,bbns3}.

\section{ results}\label{sec:results}
In the numerical calculations, parameters such as the meson mass, the Wolfenstein parameters, the
decay constants, and the lifetime of $B_s$ mesons 
are presented in Table \ref{tab:constant1}.
Other parameters relevant to the kaon-pion DAs  have been given in the second section.

\begin{table}
\caption{The decay constants of the  $J/\psi$ and $\psi(2S)$ meson are from \cite{prd90114030,epjc75293},
 while the other parameters are adopted in PDG 2016 \cite{pdg2016}
in our numerical calculations.  }
\label{tab:constant1}
\begin{tabular*}{18cm}{@{\extracolsep{\fill}}llccc}
  \hline\hline
\text{Mass(\text{GeV})}
& $M_{B}=5.28$ & $M_{B_s}=5.37$ & $m_{b}=4.66$& $m_{c}=1.275$ \\[1ex]
& $m_{\psi(2S)}=3.686$ &$m_{J/\psi(2S)}=3.097$&$m_{K}=0.494$&$m_{\pi}=0.14$    \\[1ex]
\end{tabular*}
\begin{tabular*}{18cm}{@{\extracolsep{\fill}}llccc}
 \text{The Wolfenstein parameters} &$\lambda = 0.22506$,\quad &$A=0.811$,\quad &$\bar{\rho}=0.124$,\quad &$\bar{\eta}=0.356$ \\[1ex]
\end{tabular*}
\begin{tabular*}{18cm}{@{\extracolsep{\fill}}lcccc}
\text{Decay constants (MeV)} & $f_{B}=190.9\pm 4.1$& $f_{B_s}=227.2\pm 3.4$ & $f_{\psi(2S)}=296^{+3}_{-2}$
& $f_{J/\psi}=405 \pm 14$   \\[1ex]
\end{tabular*}
\begin{tabular*}{18cm}{@{\extracolsep{\fill}}lccc}
\text{Lifetime (ps)} & $\tau_{B_s}=1.51$& $\tau_{B_0}=1.52$& $\tau_{B^+}=1.638$\\[1ex]
\hline\hline
\end{tabular*}
\end{table}

By using Eq. (\ref{eq:dfenzhibi}), integrating separately
for   the $K^*_0(1430)$ resonant and nonresonant components as well as their coherent sum,
we obtained the $CP$-averaged branching ratios for the considered  decays,
 which   are shown in Table \ref{tab:br} together
with some of the experimental measurements. 
Since the charged and neutral decay modes  differ only in the lifetimes of $B^0$ and $B^+$  in our formalism,
we can obtain the branching ratios of charged decay modes by multiplying the neutral ones by the lifetime ratio $\tau_{B^+}/\tau_{B^0}$.
Some dominant uncertainties are considered in our calculations.
The first error  is caused by the shape
parameter $\omega_b$ in the $B_{(s)}$ meson wave function. 
The second error comes from the uncertainty of the heavy quark masses.
In the evaluation, we vary the values of $m_{c(b)}$  within a $20\%$ range.
The third error is induced by  the Gegenbauer moment $B_{1}=-0.57\pm 0.13$  \cite{prd73014017,prd77014034}.
The last one is caused by the variation of the hard scale from
$0.75 t$ to $1.25 t$, which characterizes the size of the NLO QCD contributions.
The first three errors are comparable, and contribute the main uncertainties in our approach.
While the last scale-dependent uncertainty is less than $20\%$ due to the inclusion of the NLO vertex corrections.
The errors from the uncertainty of the CKM matrix elements and the decay constants of charmonia
are very small  and have been neglected.
We have checked the sensitivity of our results to the choice of the shape parameters
$a$ and $r$ [see Eq. (\ref{eq:ar})] in the LASS parametrization.
Some experimental groups \cite{prd78012004,prd72072003} prefer to choose another set of solutions with $a=2.07$ and $r=3.32$.
 Using the above values,  we find that the branching ratios  displayed in Table \ref{tab:br}  decrease  by only  a few percent.

From Table \ref{tab:br},  we find that the
$K^*_0(1430)$ resonance accounts for $41\%$ of the branching fraction and the LASS NR term accounts for $49\%$. 
 The constructive interference between them is responsible for the remaining $10\%$ in the $B\rightarrow J/\psi K\pi$ decays.
For the corresponding $\psi(2S)$ modes,
since the $K^*_0(1430)$ resonance region is very close to the upper limits of the $K\pi$ invariant mass spectra,
the resonance contribution is suppressed to $25\%$ of the total $S$-wave decay fraction.
 A similar situation also exists in the Cabbibo suppressed $B^0_s$ decay modes.
All these channels receive  a relatively large contribution from the LASS NR,
which involved the component of $\kappa$ resonance as mentioned in  \cite{prl112222002}.
In fact, the $\kappa$ fit fractions from both the  Belle \cite{prd90112009,prd88074026,prd80031104}
 and LHCb \cite{prl112222002} measurements are larger than that of  $K^*_0(1430)$ resonance.

\begin{table}
\caption{The PQCD predictions for the $CP$-averaged branching ratios
 from various components together with the $S$-wave contribution  for the considered decays. The theoretical errors correspond
to the uncertainties due to the shape parameters $\omega_b$ in the wave function of the $B_{(s)}$ meson,
 the heavy quark masses $m_b$ and $m_c$,  the Gegenbauer moment $B_1$, and  the hard scale $t$, respectively.
The experimental results are obtained  by multiplying the fit fractions by the measured  three-body branching ratios,
where all errors are combined in quadrature. }
\label{tab:br}
\begin{tabular}[t]{p{3cm}p{3cm}p{2.5cm}p{3.5cm}p{2.5cm}}
  \hline\hline
& \multicolumn{2}{c}{$B^0 \rightarrow J/\psi K^+\pi^-$ ($10^{-5}$)}&
\multicolumn{2}{c}{$B^0 \rightarrow \psi(2S) K^+\pi^-$ ($10^{-5}$)}\\
  Components & This work & Data   & This work & Data   \\
  $K^*_0(1430)$ & $6.1^{+1.4+1.1+1.5+0.4}_{-0.8-0.5-0.8-0.0}$ & $6.8^{+0.8}_{-0.6}$ \footnotemark[1]
 & $0.9^{+0.1+0.3+0.1+0.1}_{-0.1-0.1-0.1-0.0}$  & $1.7\pm 0.5$ \footnotemark[2] \\
\text{ LASS NR} & $7.2^{+2.2+1.1+1.2+1.0}_{-1.0-0.3-0.9-0.5}$ & --& $1.5^{+0.3+0.3+0.3+0.1}_{-0.3-0.2-0.2-0.1}$ & -- \\
\text{ LASS $S$-wave} & $14.8^{+3.7+2.2+3.2+1.8}_{-2.7-1.1-2.7-1.1} $ & $16.9^{+0.9}_{-0.8}$ \footnotemark[3]
& $3.4^{+0.6+0.6+0.6+0.3}_{-0.5-0.5-0.7-0.2}$ & $14.1\pm1.4$ \footnotemark[3]  \\ \hline
& \multicolumn{2}{c}{$B^0_s \rightarrow J/\psi K^-\pi^+$  ($10^{-6}$)}
 & \multicolumn{2}{c}{$B_s^0 \rightarrow \psi(2S) K^-\pi^+$ ($10^{-6}$)}\\
  Components & This work & Data   & This work & Data   \\
  $K^*_0(1430)$ & $4.1^{+0.7+0.6+1.1+0.8}_{-0.5-0.3-0.9-0.3}$ & --
  & $0.8^{+0.1+0.2+0.2+0.1}_{-0.1-0.1-0.2-0.0}$ & -- \\
\text{ LASS NR} & $4.4^{+0.6+0.6+1.0+0.6}_{-0.7-0.3-0.8-0.4}$ & --& $1.0^{+0.3+0.3+0.2+0.2}_{-0.2-0.1-0.2-0.1}$ & -- \\
\text{ LASS $S$-wave} & $8.3^{+3.7+2.2+3.2+1.8}_{-2.7-1.1-2.7-1.1} $ & --
& $2.2^{+0.6+0.6+0.4+0.4}_{-0.4-0.2-0.4-0.2}$ & $10.5\pm 2.1$ \footnotemark[4]  \\
  \hline\hline
\end{tabular}
\footnotetext[1]{The fit fractions and the measured values for $\mathcal{B}(B^0\rightarrow J/\psi K^+\pi^-)$
are given in \cite{prd90112009}.}
\footnotetext[2]{The fit fraction is obtained from a weighted average of three measurements by Belle \cite{prd80031104,prd88074026}
 and LHCb \cite{prl112222002},
 while the measured value for $\mathcal{B}(B^0\rightarrow \psi(2S) K^+\pi^-)$ is given in PDG \cite{pdg2016}.}
\footnotetext[3]{The fit fractions and the measured values for $\mathcal{B}(B^0\rightarrow (J/\psi, \psi(2S)) K^+\pi^-)$
 are given in \cite{prd79112001}.}
 \footnotetext[4]{The fit fractions and the measured values for $\mathcal{B}(B^0_s\rightarrow \psi(2S) K^-\pi^+)$
 are given in \cite{plb747484}.}
\end{table}

\begin{figure}[tbp]
\begin{center}
\setlength{\abovecaptionskip}{0pt}
\centerline{
\hspace{4cm}\subfigure{\epsfxsize=15 cm \epsffile{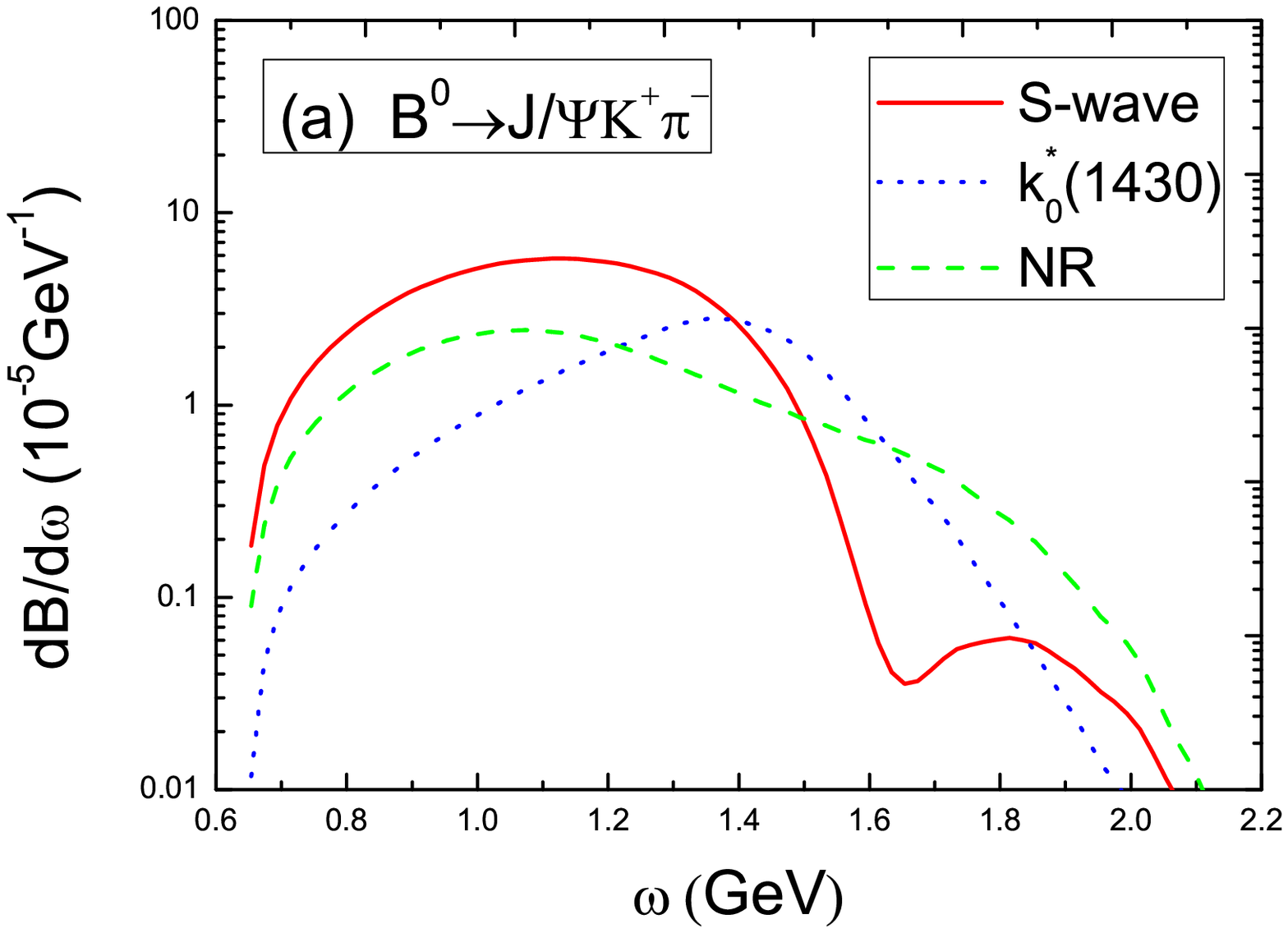} }
\hspace{-6cm}\subfigure{ \epsfxsize=15 cm \epsffile{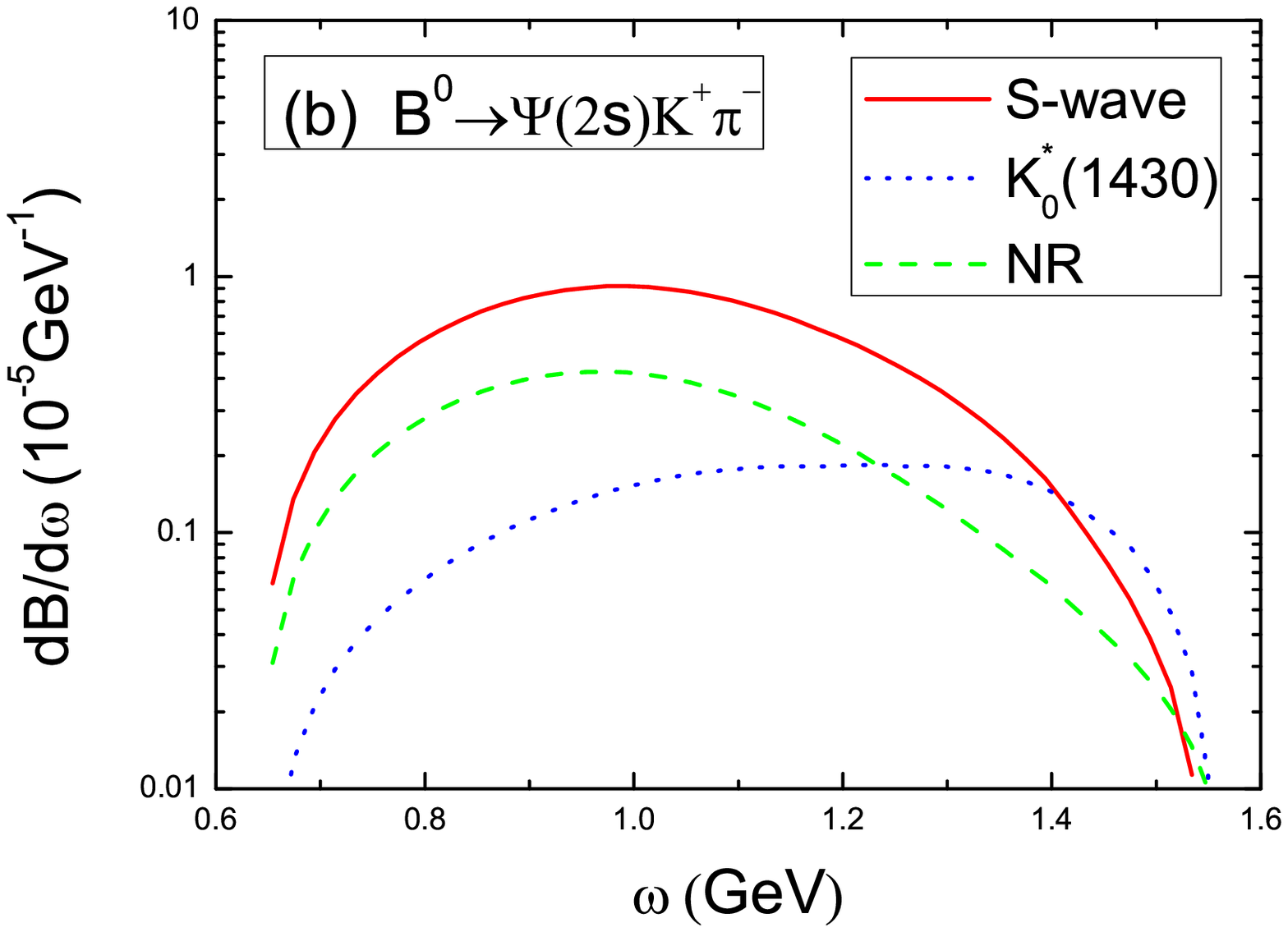}}}
\vspace{-3cm}
\centerline{
\hspace{4cm}\subfigure{\epsfxsize=15 cm \epsffile{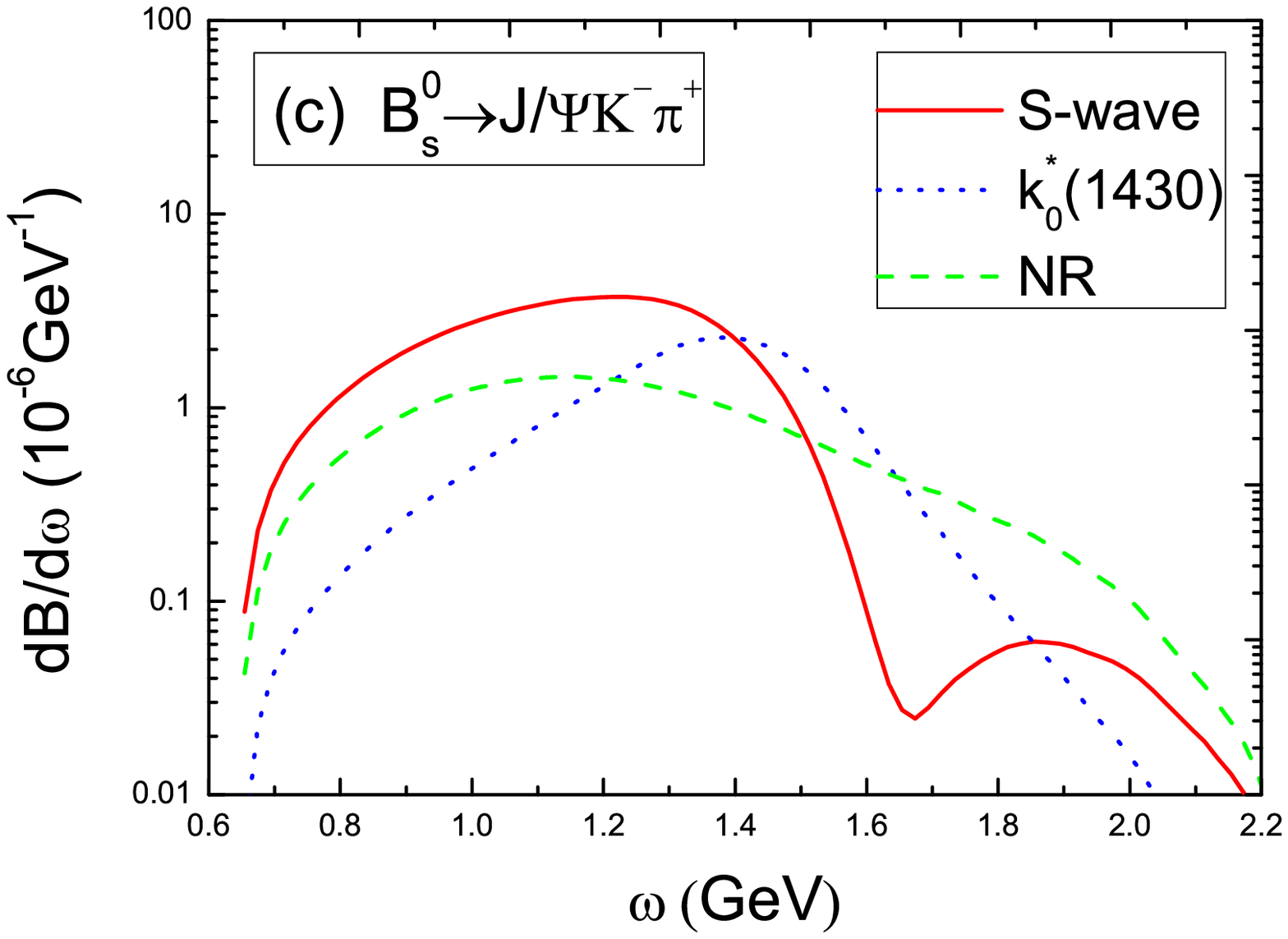} }
\hspace{-6cm}\subfigure{ \epsfxsize=15 cm \epsffile{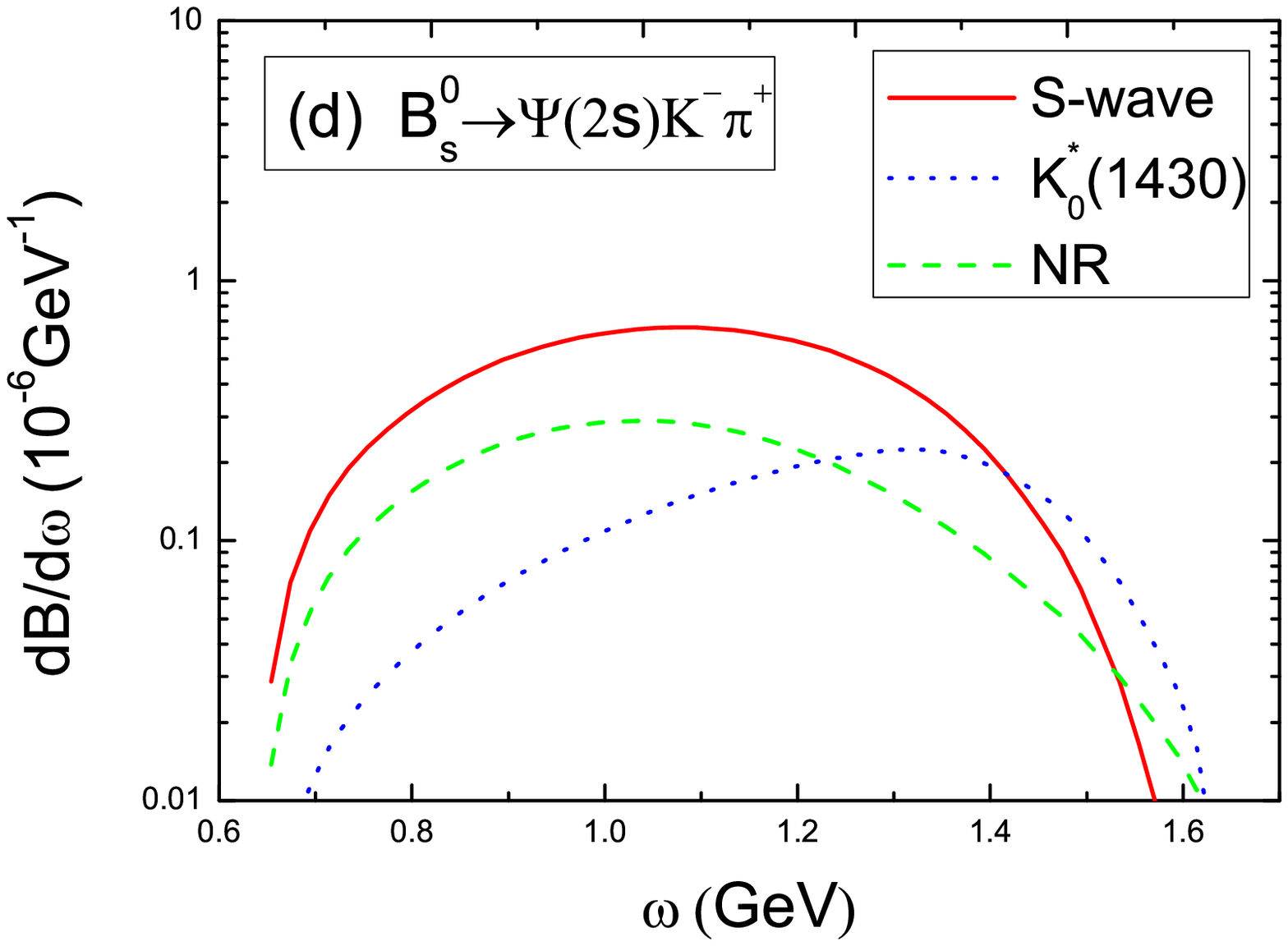}}}
\vspace{-2cm}\caption{The $\omega$ dependence of the differential decay rates $d\mathcal{B}/d\omega$ for the decay modes (a) $B^0\rightarrow J/\psi K^+\pi^-$, (b) $B^0\rightarrow \psi(2S) K^+\pi^-$,   (c) $B^0_s\rightarrow J/\psi K^-\pi^+$, and (d) $B^0_s\rightarrow \psi(2S) K^-\pi^+$ with a logarithmic $y$-axis scale. The resonance $K^*_0(1430)$ and LASS nonresonant components  are shown by the dotted blue and dashed green curves, respectively, while the solid red curves represent the  total $S$-wave contributions. }
 \label{fig:reda}
\end{center}
\end{figure}

As for the data, the fit fractions determined from the Dalitz plot analyses 
can be converted into quasi-two-body   branching fractions by multiplying
the corresponding branching fractions of the three-body decays. 
Taking the $B^0\rightarrow J/\psi K^+\pi^-$ decay as an example, based on the fit fraction of the $K^*_0(1430)$ component, which was  measured to be $f_{K^*_0(1430)}=(5.9^{+0.6}_{-0.4})\%$ with a significance of $22.0 \sigma$  by the Belle Collaboration \cite{prd90112009}, we have the center value of the quasi-two-body  branching fraction
\begin{eqnarray}
\mathcal{B}(B^0\rightarrow J/\psi K^*_0(1430)\rightarrow J/\psi K^+\pi^- )=f_{K^*_0(1430)}\times \mathcal{B}(B^0\rightarrow J/\psi K^+\pi^- ) =6.8\times 10^{-5}\;.
\end{eqnarray}
Other available fit fractions are also converted into branching fraction measurements
which are  listed in Table \ref{tab:br}.
It is shown that
the model calculations presented here are described reasonably well for the $J/\psi$ mode,
but less so for the case of  $\psi (2S)$,
especially for the $S$-wave contributions, which fall short by a large factor.
It is worth to noting that the fit fractions for the $\psi(2S)$ modes have much larger relative errors because of limited statistics.
For instance, the previous Belle Collaboration gives the fit $f_{K^*_0(1430)}=(5.3\pm 2.6)\%$ \cite{prd80031104},
while the subsequent measurements from the Belle and LHCb collaborations are $(1.1\pm 1.4)\%$  \cite{prd88074026} and $(3.6\pm 1.1)\%$ \cite{prl112222002}, respectively. Including the errors, all three measurements agree with one another.
In Table \ref{tab:br}, we calculate a weighted average and error from them  as $f_{K^*_0(1430)}=(2.9\pm 0.8)\%$, which is   closer
to the LHCb data \cite{prl112222002}.
On the other hand, comparing with the ground state $J/\psi$ modes, the branching ratio of the radially excited charmonium modes
 should be relatively small, owing to the phase space suppression and  smaller decay constants.
In Table \ref{tab:br}, our prediction of the  $S$-wave branching ratio for $B^0\rightarrow \psi(2S) K^+\pi^-$  is a few times smaller than that of
$B^0\rightarrow J/\psi K^+\pi^-$. However, the data from $BABAR$ show the same order of magnitude between the two channels.
Such a difference should be clarified  in the forthcoming experiments  based on much larger data samples.

The $B^0_s$ decay modes  can be theoretically related to the counterpart $B^0$ decays since  
 they have identical topology and similar kinematic properties in the limit of $SU(3)$ flavor symmetry.
 The relative ratios of the branching fractions for $B^0_s$ and $B^0$  decay modes
 are dominated by a Cabibbo suppression factor of $|V_{cd}|^2/|V_{cs}|^2\sim \lambda^2$ under the naive factorization approximation.
From Table \ref{tab:br}, one can see that the  $B_s$ channels have   relatively small branching ratios ($10^{-6}$).
Experimentally, the fraction of $B^0_s \rightarrow \psi(2S)K^-\pi^+$ decay proceeding via an $S$-wave is measured to be
$f_{\text{$S$-wave}}=0.339\pm 0.052$ by the LHCb experiment \cite{plb747484}, with   statistical uncertainty only.
Although the  signal $B^0_s\rightarrow J/\psi K^-\pi^+$ is found with a $4.7 \sigma$ significance  by the LHCb experiment \cite{2011025} using a mass window of $\pm 150$ MeV around the nominal $K^{*0}$ mass,
 the small size of the data sample does not permit the determination of  $K^*_0(1430)$ and the $S$-wave  fraction itself.

In Fig. \ref{fig:reda}, we plot the differential  branching ratios as functions of the $K\pi$ invariant mass $\omega$ for the
considered decays. The red (solid) curve
denotes the total $S$-wave contribution, while individual terms are given by the blue (dotted) curve
for $K^*_0(1430)$ resonance and  green (dashed) curve for LASS NR contributions.
Note that the $J/\psi-\psi(2S)$ mass difference causes significant differences in the range spanned in the respective decay modes.
As expected, 
the contributions from LASS NR   and the $K^*_0(1430)$ resonance are of comparable size. 
For the $J/\psi$ modes, the dip region near 1.6 GeV is caused by strongly destructive interference between
the resonance and  nonresonant part of the LASS parametrization.
From Eq. (\ref{eq:lass}), one can estimate that the magnitude of these two terms is approximately equal,
and the phase difference is roughly $\pi$  around the 1.6 GeV regions.
Experimentally,  it is usually interpreted as resulting from interference between the $K^*_0(1430)$
 and its  first radial excitation \cite{prd79112001}.
  However, the dip is not seen in Figs \ref{fig:reda} (b) and \ref{fig:reda} (d)
   because its region is beyond the $K\pi$ invariant mass spectra for  the $\psi(2S)$ modes.
Comparing with the $K\pi$ mass distributions obtained by $BABAR$ (Fig. 11  of Ref. \cite{prd79112001})
and LHCb (Fig. 2  of Ref. \cite{prd86071102}),
our distribution for the $S$-wave contribution 
agrees fairly well, showing a similar behavior.

\section{ conclusion}\label{sec:sum}
Motivated by the phenomenological importance of the  hadronic charmonium $B$ decays, 
in the present work we have carried out  analyses of the $B^0_{(s)}\rightarrow \psi K\pi$  decays 
within the framework of the PQCD factorization approach
by introducing the kaon-pion distribution amplitudes.
Both the $S$-wave resonant and nonresonant components
  are parametrized into the timelike scalar form factors, which can be  described by the LASS line shape.
   It is worth noting that fractions of the resonant and nonresonant components  in these decays are comparable in size.
Our predicted  $S$-wave  decay spectrum  in the kaon-pion pair invariant mass show a similar behavior as the experiment.
In particular, the $K^*_0(1430)$ production in the $B^0\rightarrow J/\psi K\pi$ decay agrees well with the results of a recent
Dalitz plot   analysis by the   Belle Collaboration.
Nevertheless, for the case of $\psi(2S)$ modes, our results for the $S$-wave branching ratios turn out to be lower than  the data.
For the $B_s$ decays, an amplitude analysis to determine the fraction of decays proceeding via an intermediate $K^*_0(1430)$
meson is still missing. We expect the relevant results could be tested by future experimental measurements.

\begin{acknowledgments}
The authors are grateful to Hsiang-nan Li
for helpful discussions. This work is supported in part by the  National Natural Science Foundation of China under
Grants No. 11547020, No. 11605060, and No. 11547038, and in part by the Program for the Top Young Innovative Talents of Higher Learning Institutions of Hebei Educational Committee under Grant No. BJ2016041.
\end{acknowledgments}

\end{document}